\documentclass[final,5p,times,twocolumn]{elsarticle}

\usepackage{graphicx}
\usepackage{xcolor}
\usepackage{subcaption}

\usepackage{amsmath, amssymb}
\usepackage{amsthm}

\usepackage[mathlines]{lineno}
\newcommand*\patchAmsMathEnvironmentForLineno[1]{%
\expandafter\let\csname old#1\expandafter\endcsname\csname #1\endcsname
\expandafter\let\csname oldend#1\expandafter\endcsname\csname
end#1\endcsname
 \renewenvironment{#1}%
   {\linenomath\csname old#1\endcsname}%
   {\csname oldend#1\endcsname\endlinenomath}%
}
\newcommand*\patchBothAmsMathEnvironmentsForLineno[1]{%
  \patchAmsMathEnvironmentForLineno{#1}%
  \patchAmsMathEnvironmentForLineno{#1*}%
}
\AtBeginDocument{%
\patchBothAmsMathEnvironmentsForLineno{equation}%
\patchBothAmsMathEnvironmentsForLineno{align}%
\patchBothAmsMathEnvironmentsForLineno{flalign}%
\patchBothAmsMathEnvironmentsForLineno{alignat}%
\patchBothAmsMathEnvironmentsForLineno{gather}%
\patchBothAmsMathEnvironmentsForLineno{multline}%
}

\usepackage{xspace}
\newcommand{\uP}{u/P\xspace}
\newcommand{\vN}{v/N\xspace}

\newcommand{\median}{\texttt{median}\xspace}

\newcommand{\sigmaSE}{\texttt{sigma-68}\xspace}
\newcommand{\mad}{\texttt{mad}\xspace}
\newcommand{\xpos}{\ensuremath{x}\xspace}
\newcommand{\zpos}{\ensuremath{m}\xspace}
\newcommand{\tpos}{\ensuremath{t}\xspace}
\newcommand{\res}{\ensuremath{R}\xspace}
\newcommand{\terr}{\ensuremath{\sigma_t}\xspace}

\newcommand{\deltares}{\ensuremath{\Delta R}\xspace}

\newcommand{\gev}{\ensuremath{\mathrm{\,Ge\kern -0.1em V}}\xspace}
\newcommand{\mev}{\ensuremath{\mathrm{\,Me\kern -0.1em V}}\xspace}
\def\mm {\ensuremath{{\,m\rm m}}\xspace}
\def\mum  {\ensuremath{{\,\mu\rm m}}\xspace}

\def\invab   {\ensuremath{\mbox{\,ab}^{-1}}\xspace}
\def\cms     {\ensuremath{{\rm \,cm}^{-2} {\rm s}^{-1}}\xspace}

\usepackage{array}
\newcommand{\PreserveBackslash}[1]{\let\temp=\\#1\let\\=\temp}
\newcolumntype{C}[1]{>{\PreserveBackslash\centering}p{#1}}
\newcolumntype{R}[1]{>{\PreserveBackslash\raggedleft}p{#1}}
\newcolumntype{L}[1]{>{\PreserveBackslash\raggedright}p{#1}}

\journal{Nuclear Instruments and Methods A}

\begin{document}

\begin{frontmatter}



\title{Measurement of the cluster position resolution of the Belle II Silicon Vertex Detector}


\author[add2211]{R.~Leboucher}
\ead{leboucher@cppm.in2p3.fr}
\author[add19]{K.~Adamczyk}
\author[add200]{L.~Aggarwal}
\author[add15]{H.~Aihara}
\author[add7]{T.~Aziz}
\author[add19]{S.~Bacher}
\author[add4]{S.~Bahinipati}
\author[add8,add9]{G.~Batignani}
\author[add222]{J.~Baudot} 
\author[add5]{P.~K.~Behera}
\author[add8,add9]{S.~Bettarini}
\author[add3]{T.~Bilka}
\author[add19]{A.~Bozek}
\author[add2]{F.~Buchsteiner}
\author[add8,add9]{G.~Casarosa}
\author[add8,add9]{L.~Corona}
\author[add12]{T.~Czank}
\author[add21]{S.~B.~Das}
\author[add222]{G.~Dujany} 
\author[add222]{C.~Finck} 
\author[add8,add9]{F.~Forti}
\author[add2]{M.~Friedl}
\author[add10,add11]{A.~Gabrielli}
\author[add10,add11]{E.~Ganiev}
\author[add11]{B.~Gobbo}
\author[add7]{S.~Halder}
\author[add16,add221]{K.~Hara}
\author[add7]{S.~Hazra}
\author[add12]{T.~Higuchi}
\author[add2]{C.~Irmler}
\author[add16,add221]{A.~Ishikawa}
\author[add17]{H.~B.~Jeon}
\author[add10,add11]{Y.~Jin}
\author[add12]{C.~Joo}
\author[add19]{M.~Kaleta}
\author[add7]{A.~B.~Kaliyar}
\author[add3]{J.~Kandra}
\author[add12]{K.~H.~Kang}
\author[add19]{P.~Kapusta}
\author[add3]{P.~Kody\v{s}}
\author[add16]{T.~Kohriki}
\author[add21]{M.~Kumar}
\author[add20]{R.~Kumar}
\author[add12]{C.~La~Licata}
\author[add21]{K.~Lalwani}
\author[add17]{S.~C.~Lee}
\author[add5]{J.~Libby}
\author[add222]{L.~Martel} 
\author[add8,add9]{L.~Massaccesi}
\author[add7]{S.~N.~Mayekar}
\author[add7]{G.~B.~Mohanty}
\author[add12]{T.~Morii}
\author[add16,add221]{K.~R.~Nakamura}
\author[add19]{Z.~Natkaniec}
\author[add15]{Y.~Onuki}
\author[add19]{W.~Ostrowicz}
\author[add8,add9]{A.~Paladino}
\author[add8,add9]{E.~Paoloni}
\author[add17]{H.~Park}
\author[add2211]{L.~Polat}
\author[add7]{K.~K.~Rao}
\author[add222]{I.~Ripp-Baudot} 
\author[add8,add9]{G.~Rizzo}
\author[add7]{D.~Sahoo}
\author[add2]{C.~Schwanda}
\author[add2211]{J.~Serrano}
\author[add16]{J.~Suzuki}
\author[add16,add221]{S.~Tanaka}
\author[add15]{H.~Tanigawa}
\author[add2]{R.~Thalmeier}
\author[add7]{R.~Tiwary}
\author[add16,add221]{T.~Tsuboyama}
\author[add15]{Y.~Uematsu}
\author[add19]{O.~Verbycka}
\author[add10,add11]{L.~Vitale}
\author[add15]{K.~Wan}
\author[add15]{Z.~Wang}
\author[add1]{J.~Webb}
\author[add9]{J.~Wiechczynski}
\author[add2]{H.~Yin}
\author[add2211]{L.~Zani}
\author[]{\\ \vspace{1 mm} (Belle-II SVD Collaboration)}
\address[add1]{School of Physics, University of Melbourne, Melbourne, Victoria 3010, Australia}
\address[add2]{Institute of High Energy Physics, Austrian Academy of Sciences, 1050 Vienna, Austria}
\address[add3]{Faculty of Mathematics and Physics, Charles University, 121 16 Prague, Czech Republic}
\address[add2211]{Aix Marseille Universit$\acute{e}$ , CNRS/IN2P3, CPPM, 13288 Marseille, France}
\address[add222]{IPHC, UMR 7178, Universit$\acute{e}$ de Strasbourg, CNRS, 67037 Strasbourg, France} 
\address[add4]{Indian Institute of Technology Bhubaneswar, Satya Nagar, India}
\address[add5]{Indian Institute of Technology Madras, Chennai 600036, India}
\address[add21]{Malaviya National Institute of Technology Jaipur, Jaipur 302017, India} 
\address[add20]{Punjab Agricultural University, Ludhiana 141004, India} 
\address[add200]{Panjab University, Chandigarh 160014, India} 
\address[add7]{Tata Institute of Fundamental Research, Mumbai 400005, India}
\address[add8]{Dipartimento di Fisica, Universit\`{a} di Pisa, I-56127 Pisa, Italy}
\address[add9]{INFN Sezione di Pisa, I-56127 Pisa, Italy}
\address[add10]{Dipartimento di Fisica, Universit\`{a} di Trieste, I-34127 Trieste, Italy}
\address[add11]{INFN Sezione di Trieste, I-34127 Trieste, Italy}
\address[add221]{The Graduate University for Advanced Studies (SOKENDAI), Hayama 240-0193, Japan} 
\address[add12]{Kavli Institute for the Physics and Mathematics of the Universe (WPI), University of Tokyo, Kashiwa 277-8583, Japan}
\address[add15]{Department of Physics, University of Tokyo, Tokyo 113-0033, Japan} 
\address[add16]{High Energy Accelerator Research Organization (KEK), Tsukuba 305-0801, Japan}
\address[add17]{Department of Physics, Kyungpook National University, Daegu 41566, Korea}
\address[add19]{H. Niewodniczanski Institute of Nuclear Physics, Krakow 31-342, Poland}

\begin{abstract}
The Silicon Vertex Detector (SVD), with its four double-sided silicon strip sensor layers, is one of the two vertex sub-detectors of Belle II operating at SuperKEKB collider (KEK, Japan). Since 2019 and the start of the data taking, the SVD has demonstrated a reliable and highly efficient operation, even running in an environment with harsh beam backgrounds that are induced by the world's highest instantaneous luminosity.


In order to provide the best quality track reconstruction with an efficient pattern recognition and track fit, and to correctly propagate the uncertainty on the hit’s position to the track parameters, it is crucial to precisely estimate the resolution of the cluster position measurement. Several methods for estimating the position resolution directly from the data will be discussed.

\end{abstract}


\end{frontmatter}

\section*{Introduction}
\label{sec: Introduction}
The Belle II experiment~\cite{B2TechnicalDesignReport} operates at the high-energy physics intensity frontier and searches for physics beyond the Standard Model~\cite{B2PhyBook} in rare \(b\), charm and tau decays. Belle II is collecting data since March 2019 at the \(e^+e^-\) asymmetric energy collider SuperKEKB~\cite{SuperKEKBDesign} at Tsukuba, in Japan, mainly at the centre of mass energy of the \(\Upsilon(4S)\) resonance, \(10.58\gev\). By achieving a design instantaneous luminosity of \(6\times10^{35}\cms\), it will collect a final data set of \(50\invab\). Relevant features of the Belle II detector to attain its main physics goals are the precise  and efficient track reconstruction capabilities, including for those with low momentum, decay vertex determination and identification of different kinds of charged particles. The vertex detector plays a crucial role in fulfilling each of these requirements. It is composed of two layers of DEPFET pixel sensors (PXD), with the innermost layer at 1.4 cm from the interaction point, and 4 layers of double-sided silicon strip sensors (SVD).

\section{The Belle II silicon strip detector}
\label{sec: Belle II silicon strip detector}
The Silicon Vertex Detector SVD~\cite{B2SVDVertexYuma} is composed of 172 double-sided silicon strip detectors (DSSD) distributed in four layers of 7, 10, 12 and 16 ladders with 2, 3, 4 and 5 sensors each, for a total material budget of  0.7\% of the radiation length per layer on average. Along the sensors, strips are arranged in perpendicular directions on opposite sides and the signal is collected by the APV25 chips which provide an analog readout: the \uP side measures the \(r\phi\)-direction and the \vN side provides information on the \(z\)-coordinate along the beam line.

\begin{table}[htbp]
    \centering
    \begin{tabular}{L{3.5cm}||C{0.95cm}C{0.95cm}C{1.8cm}}
        {} & Small & Large & Trap. \\
        \hline\hline
        No. \uP readout strips & 768 & 768 & 768 \\
        No. \vN readout strips & 768 & 512 & 512 \\
        Readout pitch \uP strips (\mum) & 50 & 75 & 50-75 \\
        Readout pitch \vN strips (\mum) & 160 & 240 & 240 \\
        Sensor thickness (\mum) & 320 & 320 & 300 \\
        Active Length (\mm) & 122.90 & 122.90 & 122.76 \\
        Active Width (\mm) & 38.55 & 57.72 & 57.59-38.42 \\
    \end{tabular}
    \caption{Geometrical details of the SVD DSSD sensors. All sensors have one intermediate floating strip between two readout strips.}
    \label{tab: Geometrical DSSD sensors}
\end{table}
Layer 3 is equipped with “small” rectangular sensors, Layer 4-5-6 are build with “large” rectangular sensors and a “trapezoidal” one. The geometrical details of the different sensors are shown in Tab.~\ref{tab: Geometrical DSSD sensors}.

The SVD plays a crucial role in reconstructing the decay vertex and low-momentum particles, providing stand-alone tracking capabilities and contributing to charged particle identification through the ionisation energy-loss information. Moreover, it contributes to extrapolating the tracks towards the PXD and defining the region of interests to reduce the PXD data size.
An excellent cluster position resolution is mandatory for SVD reconstruction, and it is a crucial input for tracking to improve the quality of reconstructed tracks and vertices, moreover its knowledge is necessary to correctly propagate the uncertainty on the track extrapolation.

\section{Cluster position resolution analysis strategy}
\label{sec: Methods}
\begin{figure}[!htbp]
    \centering
    \includegraphics[width=0.65\columnwidth,keepaspectratio]{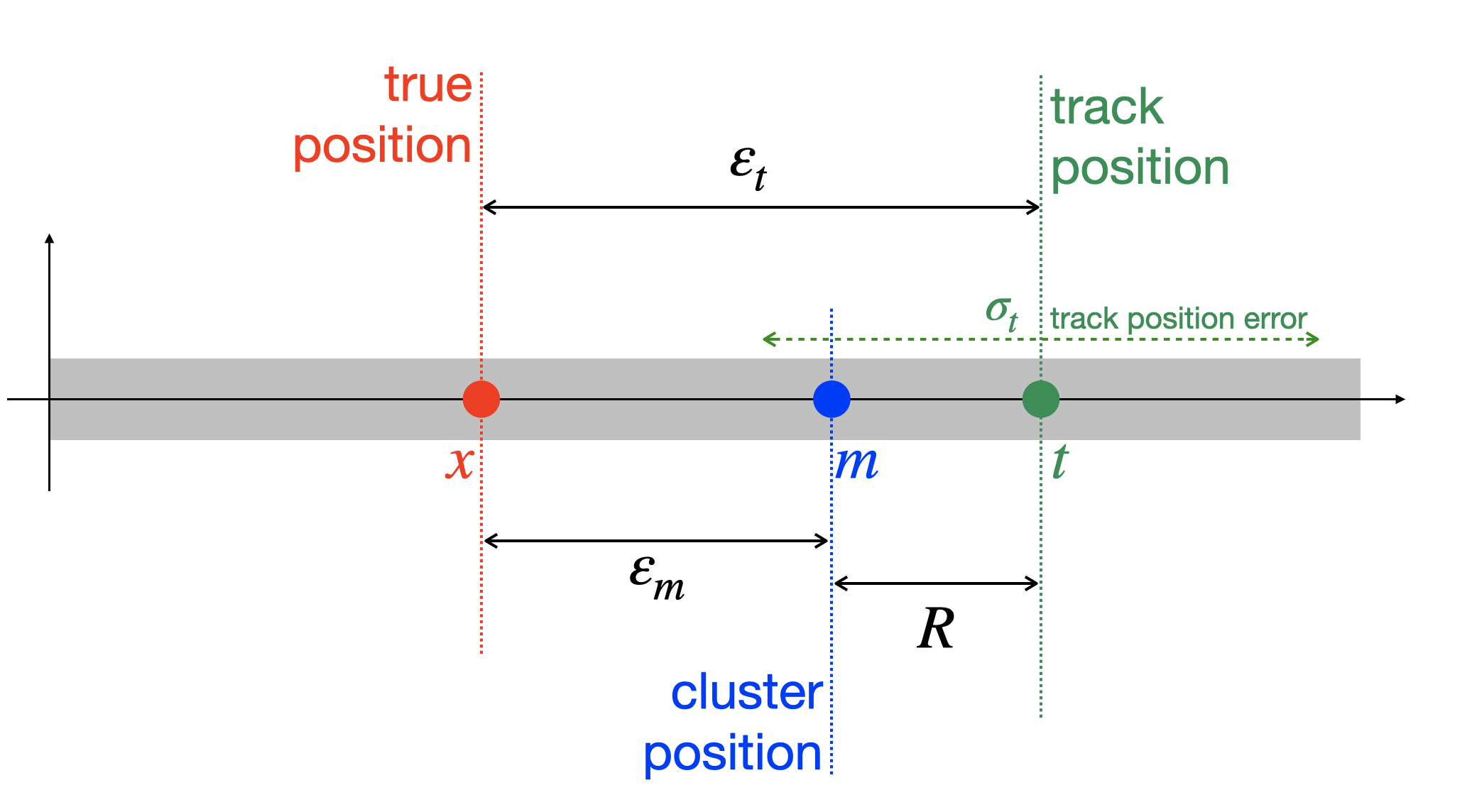}
    \caption{Schematic view of a sensor plane with the cluster position \zpos, the \emph{unbiased} track position \tpos and the true position \xpos only available in simulations, the measured residuals \res.}
    \label{fig: Positions ingredients}
\end{figure}

The tracks traversing the SVD sensors activate adjacent strips that are gathered into clusters. To each reconstructed cluster we assign (Fig.~\ref{fig: Positions ingredients}): 
\begin{itemize}
    \item the cluster position \(\zpos=\frac{\sum_{i} X_iS_i}{\sum_{i} S_i}\) obtained as the center of gravity of all strips position \(X_i\) of the given cluster weighted by the charge collected on each strip \(S_i\); 
    \item an \emph{unbiased} track position intercept \tpos from the track finding~\cite{B2TrackFinding} and its error \terr, where the track reconstruction is performed excluding the cluster considered for the resolution measurement;
    \item the true position \xpos, only in simulation.
\end{itemize}
The cluster position resolution is extracted from the residuals $\res=\zpos-\tpos$ of the cluster position with respect to the unbiased track intercept position and the effect of the track extrapolation error is subtracted.

We consider three distinct approaches in this paper:
\begin{itemize}
    \item \textbf{Event by Event} (EBE): consists in removing event-by-event the error on track extrapolation \terr from the residual \res in quadrature. 
    \begin{equation}
        \sigma_{cl}^{\rm EBE} = \sqrt{\langle R^2-\terr^2\rangle}_{trunc}.
    \end{equation}
    Here, \(trunc\) refers to the truncation of \(\res^2-\terr^2\) optimized on the simulation to match the true resolution, defined as the width of the distribution of \(\zpos-\xpos\). The truncation is needed to eliminate the long non-Gaussian tail of the \(\res^2-\terr^2\) distribution.

\item \textbf{Global}: differently from the EBE method, the so-called "global method" aims at removing the contribution of \terr by subtracting in quadrature from the width of the residuals the width and central value of the track error. The resolution is finally extracted by:
    \begin{equation}\label{eq: Global Method}
        \sigma_{cl}^{\rm GL} = \sqrt{\mad^2(R)-\median^2(\terr)-\mad^2(\terr)}.
    \end{equation}
    The median is used as estimator of the central value of \terr distribution as is is robust against the outliers. For the same reason the widths of \res and \terr distribution are estimated with the \textit{median absolute deviation}, which is defined, for variable \(y\), as \(\mad(y)=1.4826\times\median(\left|y-\median(y)\right|)\).

    \begin{figure}[!htbp]
        \centering
        \includegraphics[width=0.45\columnwidth,keepaspectratio]{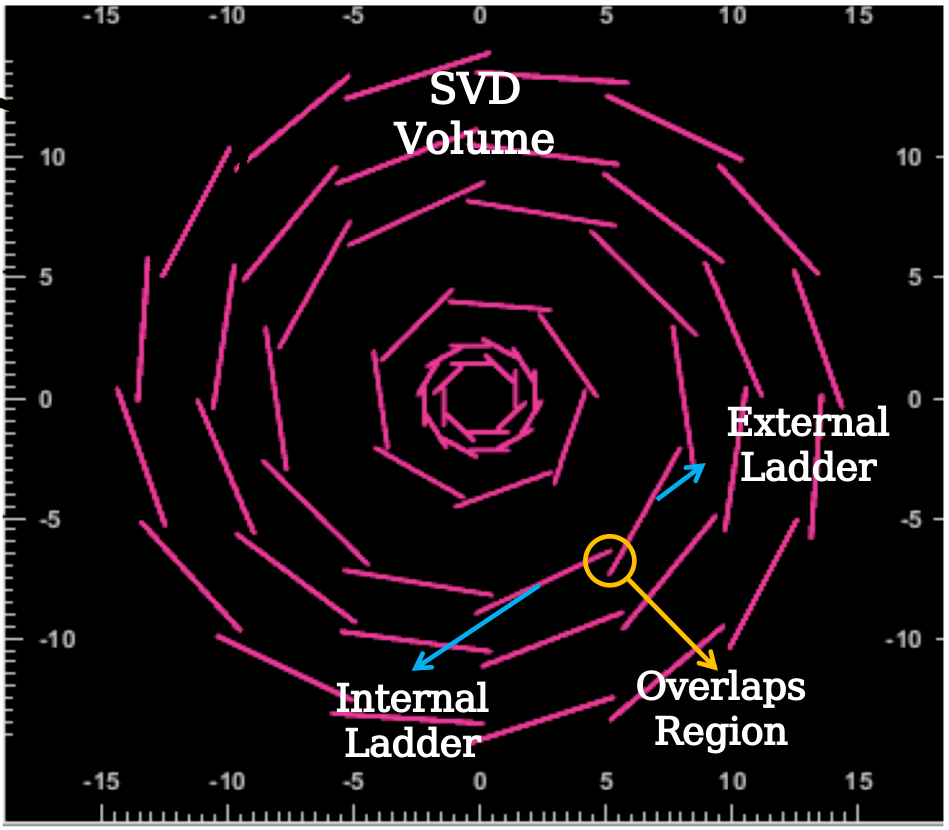}
        \caption{Schematic view of the SVD volume in the \(r\phi\) direction.}
        \label{fig: Scheme overlpas region}
    \end{figure}
\item \textbf{Pair-method}: an alternative strategy~\cite{cms_tracker_collaboration_stand_alone_2009} is implemented in Belle II thanks to the SVD's windmill architecture, Fig.~\ref{fig: Scheme overlpas region}.
    The tracks are reconstructed, accepting only those with two hits in the same layer and on consecutive ladders and in the fiducial area. The residuals \(\zpos-\tpos\) determined on both overlapping ladders (internal and externals) are then subtracted to define the double residual \deltares. The double residuals are geometrically corrected to account for the non-parallel sensors on a same layer. Then \deltares is fitted with a Student's t-distribution \(T\) with the parameters: the number of degrees of freedom \(\nu\); the mean of the distribution \(\mu\), and the variance \(\sigma^{2}\). The resolution \(\sigma_{cl}^{\rm pair}\) is finally defined as the width of \(T\) computed as the \sigmaSE~\footnote{\sigmaSE is half distance between 16th and 84th quantiles.}:
    \begin{equation}
        \sigma_{cl}^{\rm pair} = \sigmaSE\left(T(X,\nu,\mu,\sigma)\right).
    \end{equation}

    \begin{table}[htbp]
        \centering
        \begin{tabular}{l||C{1.2cm}C{1.2cm}C{1.2cm}C{1.2cm}}
            {} & L3 & L4 & L5 & L6 \\
            \hline\hline
            Internal & $20^{\circ}$;$25^{\circ}$ & $5^{\circ}$;$15^{\circ}$ & $5^{\circ}$;$10^{\circ}$ & $0^{\circ}$;$5^{\circ}$ \\
            External & $-35^{\circ}$;$-30^{\circ}$ & $-30^{\circ}$;$-20^{\circ}$ & $-25^{\circ}$;$-20^{\circ}$ & $-25^{\circ}$;$-15^{\circ}$
        \end{tabular}
        \caption{Incident angle acceptance region for the \uP side in the pair-method.}
        \label{tab: OverlapsAngularacceptanceuP}
    \end{table}
    The double residual has the advantage of being independent from the error on the extrapolated track intercept position, which is cancelled out in the double subtraction. The method decouples the contribution of the tracking precision from the actual cluster position resolution and is only marginally sensitive to the Coulomb scattering thanks to the small radial distance between the two overlapped sensors, however the incident angular range is limited following the \uP side as shown in Tab.~\ref{tab: OverlapsAngularacceptanceuP}.
\end{itemize}

\section{SVD resolutions measurement}
\label{sec: SVD Resolutions}
\begin{figure*}[!ht]
    \centering
    \begin{subfigure}[b]{0.45\textwidth}
       \centering 
       \includegraphics[width=\textwidth,keepaspectratio]{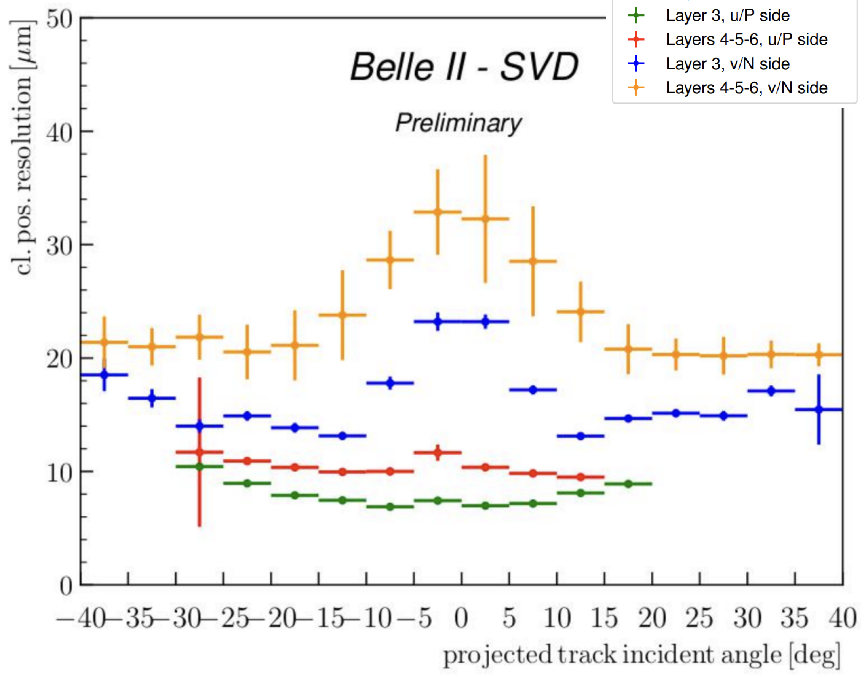}
       \includegraphics[width=\textwidth,keepaspectratio]{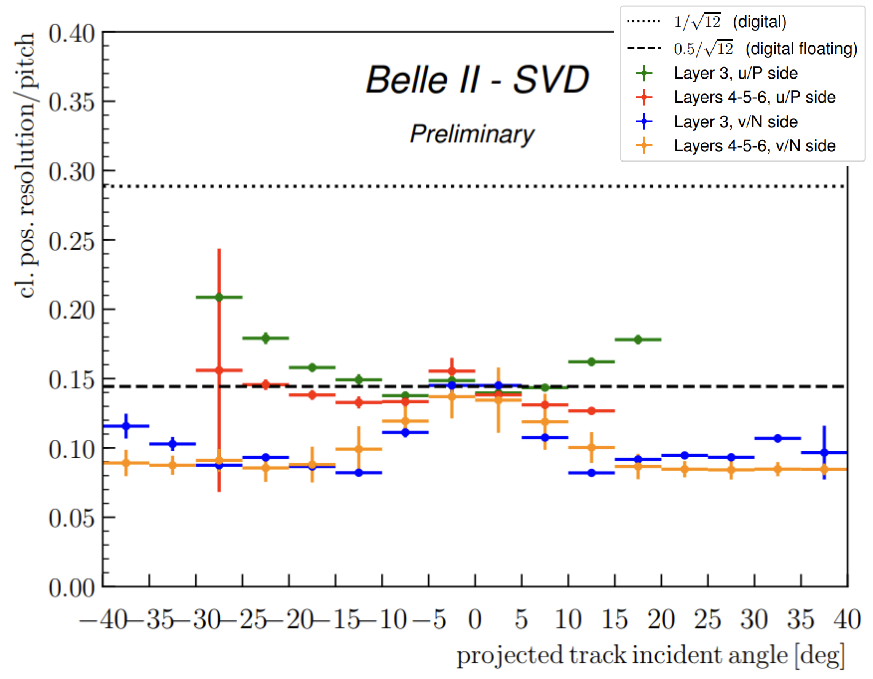}
       \caption{Determined with the Event by Event method}
       \label{fig: Resolution results for EBE}
    \end{subfigure}
    ~~
    \begin{subfigure}[b]{0.45\textwidth}
        \centering
        \includegraphics[width=\textwidth,keepaspectratio]{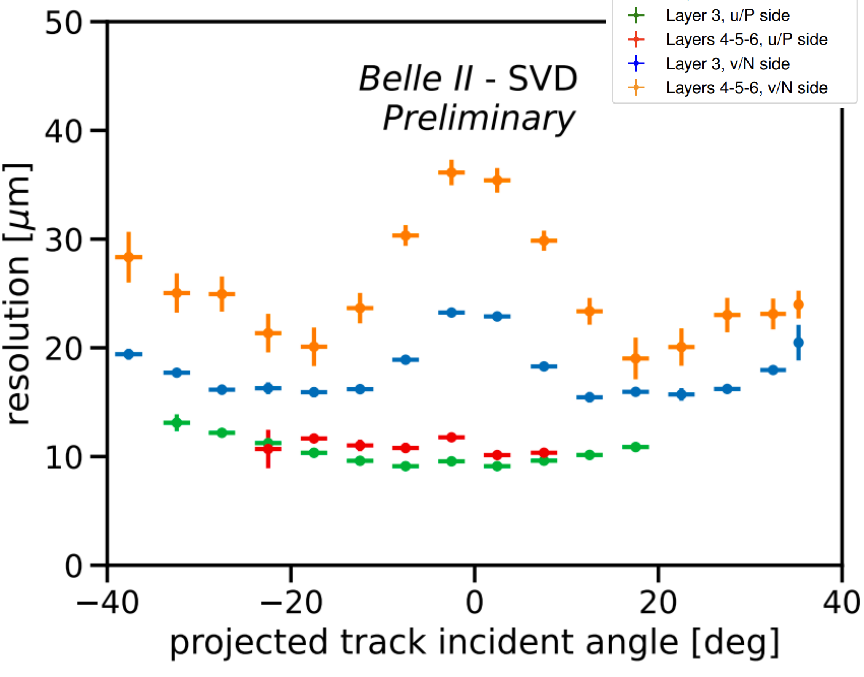}
        \includegraphics[width=\textwidth,keepaspectratio]{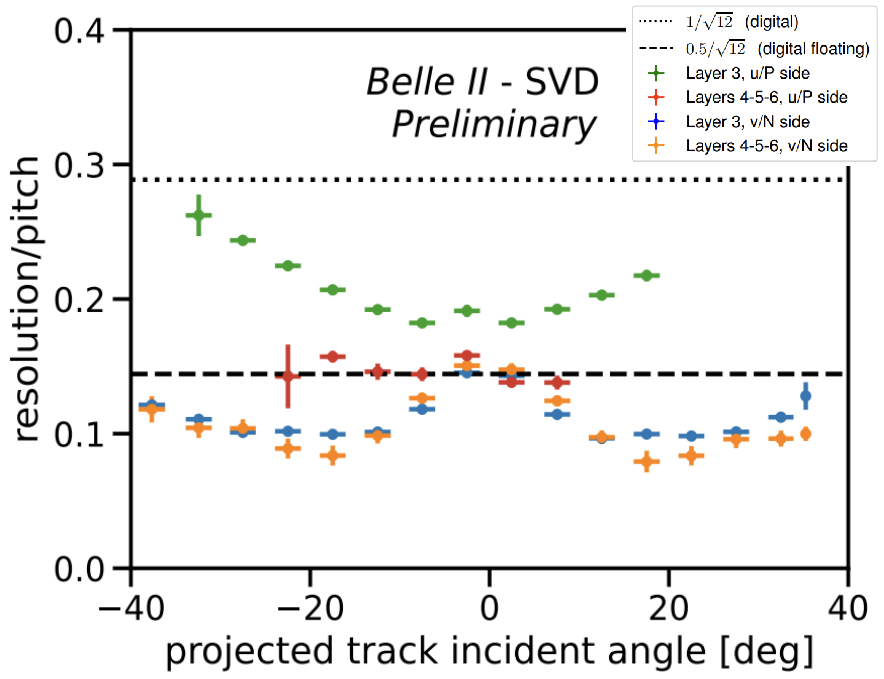}
        \caption{Determined with the Global method}
        \label{fig: Resolution results for Global}
    \end{subfigure}
    \caption{Cluster position resolution (top) and resolution over strip pitch (bottom) for each sides and layers following the track incident angle.}
    \label{fig: Resolution results}
\end{figure*}
The cluster position resolution is measured using \(e^+e^- \rightarrow \mu^+\mu^-\) data, as a function of the track incident angle in bins from \(-40^{\circ}\) to \(40^{\circ}\) with a \(5^{\circ}\) step. Results are shown for the EBE method in Fig.~\ref{fig: Resolution results for EBE} and for the global method in Fig.~\ref{fig: Resolution results for Global}. In both methods, the measured resolutions has the expected shape, showing a minimum at the incident angle for which the projection of the track along the direction perpendicular to the strips on the detector plane corresponds to two strip pitches. Given the various sensor pitches with one floating strip, the minimum is expected at $4^{\circ}$ ($7^{\circ}$) for the \uP side and at $14^{\circ}$ ($21^{\circ}$) on the \vN side, for layer 3 (4, 5 and 6).

In addition, the measured resolution for normal incident tracks is in fair agreement with digital resolution, that with a floating strip is equal to \(pitch/(2\sqrt{12})\), with 11\mum for u/P layer 4, 5 and 6; and 25 (35) for v/N layer 3 (4, 5 and 6) as shown in Fig.~\ref{fig: Resolution results for EBE} and Fig.~\ref{fig: Resolution results for Global}. The digital resolution provides a reference value only under the assumption that a track activates only one strip. Tracks with larger incident angle activate more strips and indeed our resolution is better than the digital one thanks to the analog readout. Layer 3 u/P side, the sensors with the smallest pitch, represents an exception: the measured resolution is slightly worse than the digital one for perpendicular tracks and it degrades with larger angles. This is effect is still under investigation.

\begin{figure*}[!ht]
    \centering
    \includegraphics[width=.45\textwidth,keepaspectratio]{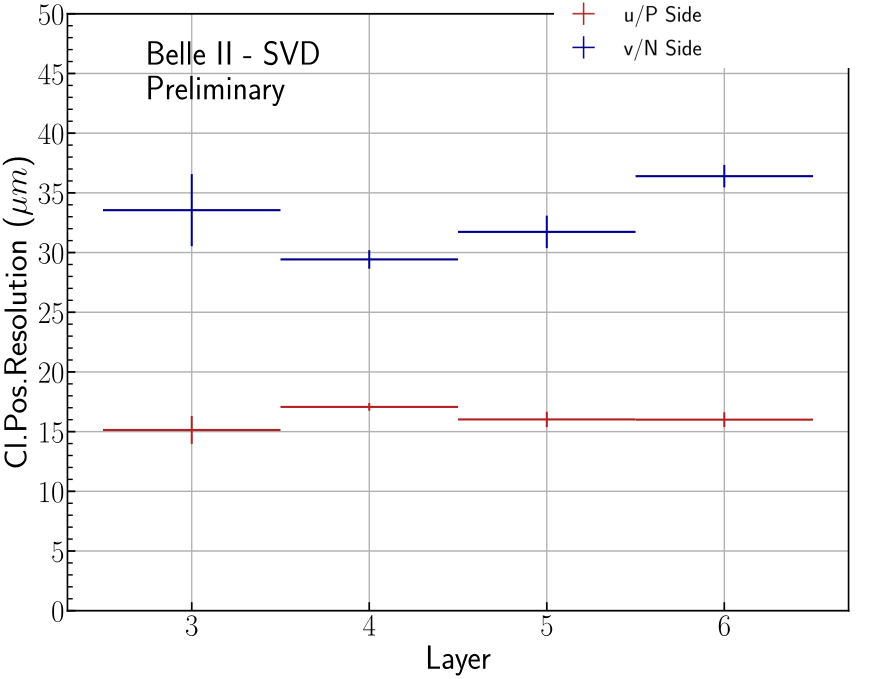}
    \includegraphics[width=.45\textwidth,keepaspectratio]{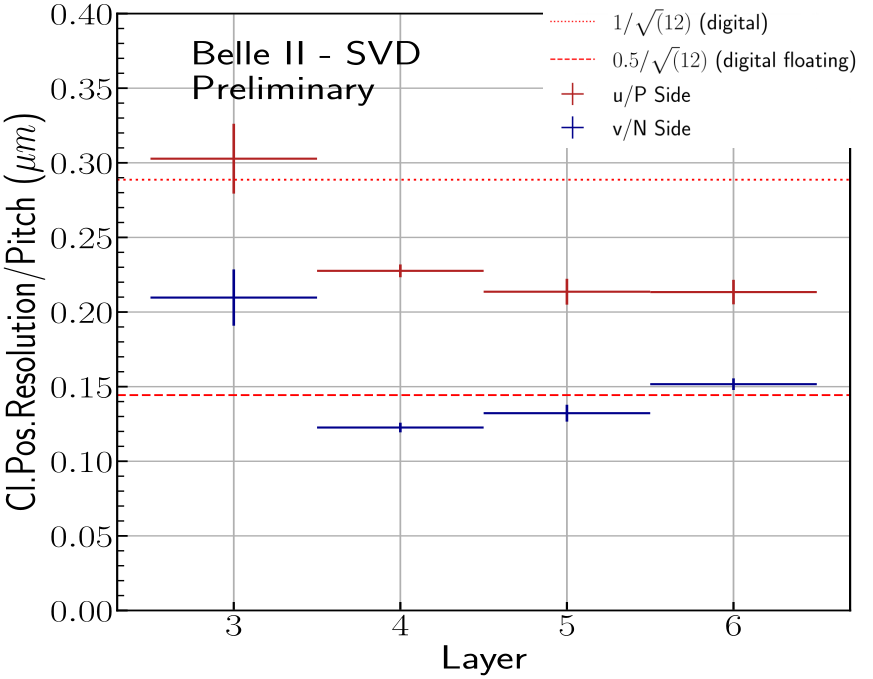}
    \caption{Cluster position resolution (top) and resolution over strip pitch (bottom) determined by the pair method for each sides and layers following the track incident angle.}
    \label{fig: Resolution results for Overlap}
\end{figure*}

Differently from EBE and Global methods, the pair method is sensitive to a limited incident angle range in \uP side by construction of the overlapping region in the \(r\phi\) direction, therefore the resolutions measurement reported in Fig.~\ref{fig: Resolution results for Overlap} is averaged on the whole accessible angular range.
The resolution measured with the pair method shown in Fig.~\ref{fig: Resolution results for Overlap} and detailed in Tab.~\ref{tab: Measuredresolution} are higher than the measured resolution with the other methods, except for the \vN outermost layers. This behaviour is not fully understood, but will be the subject of future studies. 

\section*{Conclusions and Outlooks}
\label{sec: Conclusions and Outlooks}
Since the start of the data taking the SVD has demonstrated a reliable and highly efficient operation, with excellent performances confirmed by the measurement of the cluster position resolution, summarized Tab.~\ref{tab: Measuredresolution}. Some studies to investigate the small discrepancies observed on layer 3 \uP side and to further investigate larger deviation measured on pair overlaps method are planned.

\begin{table}[htbp]
    \centering
    \begin{tabular}{l||cccc}
         {} & Digital & EBE & Global & Pair  \\
         \hline\hline
         Layer 3 \uP (\mum) & 7 & 7 & 9 & 15 \\
         Layer 456 \uP (\mum) & 11 & 10 & 11 & 16-17 \\
         Layer 3 \vN (\mum) & 23 & 24 & 23 & 33 \\
         Layer 456 \vN (\mum) & 35 & 32 & 35 & 29-36 \\
    \end{tabular}
    \caption{Summary of the digital and measured resolution taken at the normal incidence for the EBE and Global methods. For the Pair method the average  on the whole accessible angular range is shown.}
    \label{tab: Measuredresolution}
\end{table}

\section*{Acknowledgements}
This project has received funding from the European Union's Horizon 2020 research and innovation programme under the Marie Sklodowska-Curie grant agreements No 644294 and 822070 and ERC grant agreement No 819127. This work is supported by MEXT, WPI, and JSPS (Japan); ARC (Australia); BMBWF (Austria); MSMT (Czechia); CNRS/IN2P3 (France); AIDA-2020 (Germany); DAE and DST (India); INFN (Italy); NRF and RSRI (Korea); and MNiSW (Poland).



\bibliographystyle{elsarticle-num}
\bibliography{bibliography.bib}







\end{document}